\documentstyle[12pt]{article} \textwidth=6truein \textheight=21truecm
\begin{document} \thispagestyle{empty}

\noindent\hfill  OHSTPY-HEP-TH-92-023\\
\begin{center}\begin{Large}\begin{bf}
Spontaneous Symmetry Breaking of $\phi^4_{1+1}$ in Light-Front Field Theory\\
\end{bf}\end{Large}\end{center}
\vspace{.75cm}\begin{center}
Carl M. Bender\\[10pt]
      \end{center}
    \vspace{0.1cm}
     \begin{center}
     \begin{it}
Department of Physics\\
Washington University\\
St. Louis, MO 63130\\
       \end{it}
        \end{center}
        \vspace{0.5cm}
      \begin{center}
Stephen Pinsky and Brett Van de Sande\\[10pt]
     \end{center}
      \vspace{0.1cm}
      \begin{center}
      \begin{it}
Department of Physics\\
The Ohio State University\\
174 West 18th Avenue\\
Columbus, Ohio  43210\\
       \end{it}
         \end{center}
\vspace{1cm} \baselineskip=35pt \begin{abstract} \noindent
We study spontaneous symmetry breaking in $(\phi^4)_{1+1}$ using the
light-front formulation of the field theory. Since the physical vacuum is
always
the same as the perturbative vacuum in light-front field theory the fields must
develop a vacuum expectation value through the zero-mode components of the
field. We solve the nonlinear operator equation for the zero-mode in the
one-mode approximation. We find that spontaneous symmetry breaking occurs at
$\lambda_{\rm critical} = 4\pi\left(3+\sqrt 3\right)$, which is consistent with
the value $\lambda_{\rm critical} = 54.27$ obtained in the equal time theory.
We
calculate the value of the vacuum expectation value as a function of the
coupling constant in the broken phase both numerically and analytically using
the $\delta$ expansion. We find two equivalent broken phases. Finally we show
that the energy levels of the system have the expected behavior within the
broken phase.
\end{abstract}
\newpage

\begin{large} \begin{bf} \noindent I. Introduction   \end{bf} \end{large}
\vspace{.1in}
\baselineskip=18pt

It was Dirac \cite{dirac} who first recognized that different generators of the
Poincar\'{e} group could be used as a Hamiltonian for the purpose of quantizing
a field theory. He showed that, among these, light-front quantization was
unique. In light-front quantization the Hamiltonian is $p^- = \left(p^0 -
p^z\right)/\sqrt 2$, and $p^+ = \left(p^0 + p^z\right)/\sqrt 2$ is the third or
``longitudinal'' component of momentum. The longitudinal momentum has a
positive
semi-definite spectrum, and massive excitations cannot mix with the vacuum.
Therefore, the bare Fock space vacuum is an eigenstate of the full Hamiltonian.
Since theories such as QCD exhibit spontaneous symmetry breaking, spontaneous
symmetry breaking must appear through some other mechanism in the context
of light-front field theory. We will see that spontaneous symmetry breaking
occurs in light-front theory because the field includes a zero-mode that is not
an independent degree of freedom. This mode is a complicated operator-valued
function of all other modes in the theory and may lead to a non-zero vacuum
expectation value (VEV).

In order to investigate the zero-mode in light-front field theory, we will
consider a discretized $\phi^4$ field theory in two space-time dimensions.
The formal constraint equation obtained from the Dirac-Bergmann quantization
procedure relates the zero-modes to all the other modes in the problem
\cite{wittman}. This equation is, however, most easily obtained by integrating
the equation of motion \cite{robertson}. It is clearly very difficult to solve
and, to date, only approximate qualitative solutions have been generated
\cite{heinzl}. The procedure that generates the constraint equation does not
specify an operator ordering. We will use a symmetric operator-ordering
prescription \cite{benderpinsky}. In order to render the problem tractable, we
will truncate the Fock space to include only the lowest-energy mode. We will
see
that for weak coupling the theory has only the trivial solution for the
zero-mode. However, as we increase the coupling we reach a critical coupling
where a
pair of nontrivial  solutions to the constraint equation appear.

Section II presents a simple derivation of the zero-mode constraint equation in
the classical case. Section III briefly discusses quantization and mass
renormalization of the theory. In Section~IV we discuss the asymptotic behavior
of
the zero-mode in the large-particle-number sectors. From the asymptotic
behavior
of the constraint equation we show that the theory has a critical coupling. In
Section V we use the $\delta$ expansion to study the solution branches away
from the
critical point. Section VI present some numerical solutions to the equation. We
show that the theory has a critical point predicted by the asymptotic behavior
of the constraint equation. The two nontrivial solutions are plotted as
functions of the coupling and the VEV along with the $\delta$ expansion
solution. We also study the behavior of matrix elements of the zero-mode near
the critical curves and the energy eigenvalue as a function of the coupling.
Finally in Section VI we discuss our results and the remaining work that is
needed
on this problem.\\

\begin{large}\begin{bf}\noindent
II.  The Classical Case
\end{bf}\end{large}

\vspace{.1in}
The details of the Dirac-Bergmann prescription and its application to the
system considered in this paper are discussed elsewhere in the literature
\cite{wittman,heinzl}. In terms of light-front coordinates $x^\pm = \left(x^0
\pm x^1\right)/\sqrt{2}$.  For a classical field the $(\phi^4)_{1+1}$
Lagrangian
is

$${\cal L} = \partial_+\phi\partial_-\phi - {{\mu^2}\over 2} \phi^2 - {\lambda
\over 4!} \phi^4\;.\eqno(2.1)$$

\vspace{.1in}
\noindent We put the system in a box of length $d$ and impose periodic boundary
conditions. For most of our discussion we work in momentum space. We define
$q_k$ by

$$\phi\!\left(x\right) = {1\over{\sqrt d}} \sum_n q_n(x^+) e^{i
k_n^+ x^-}\; ,\eqno(2.2)$$

\vspace{.1in}
\noindent where $k_n^+ =2\pi n/d$ and the summations run over all integers
unless otherwise noted.

Next, we introduce some notation and separate out the zero-mode. We define

$$\Sigma_n = {1\over{n!}} \sum_{i_1, i_2, \ldots, i_n \neq 0} q_{i_1} q_{i_2}
\ldots q_{i_n}\, \delta_{i_1 + i_2 + \ldots + i_n, 0}\;.\eqno(2.3)$$

\vspace{.1in}
\noindent Using the Dirac-Bergmann prescription, one can find the canonical
Hamiltonian
$$P^- = {{\mu^2 q_0^2}\over 2} + \mu^2 \Sigma_2 + {{\lambda q_0^4}\over{4! d}}
+
{{\lambda q_0^2 \Sigma_2}\over{2! d}} + {{\lambda q_0 \Sigma_3}\over d} +
{{\lambda \Sigma_4}\over d}\;.\eqno(2.4)$$

\vspace{.1in}
\noindent Following the Dirac-Bergmann prescription, we identify first-class
constraints which define the conjugate momenta
$$0 = p_n - i k_n^+ q_{-n}\;,\eqno(2.5)$$

\vspace{.1in}
\noindent and a secondary constraint,
$$0 = \mu^2 q_0 + {{\lambda q_0^3}\over{3! d}} + {{\lambda q_0 \Sigma_2}\over
d} + {{\lambda \Sigma_3}\over d}\; ,\eqno(2.6)$$

\vspace{.1in}
\noindent which determines the ``zero-mode'' $q_0$. This result can also be
obtained by integrating the equations of motion. One can calculate the
Dirac brackets between the coordinates $q_n$,
$$\left[q_m,q_n\right] = {d\over{4 \pi i m}}\, \delta_{m+n, 0}\;,\quad m, n
\neq 0\; ,\eqno(2.7)$$
$$\left[q_0,q_n\right] = {\lambda\over{4 \pi i m}} {{d
\displaystyle\sum_{k, l} q_k q_l \delta_{k+l, n}}\over{2 d \mu^2 + \lambda
{\displaystyle\sum_m q_m q_{-m}}}} \; ,\quad n \neq 0\eqno(2.8)$$

\vspace{.1in}
\noindent and with the conjugate momenta,
$$\left[q_m,p_n\right] = {{\delta_{n, m}}\over 2}\;,\quad m ,n \neq
0\;.\eqno(2.9)$$

\vspace{.1in}
\noindent The total longitudinal momentum is given by
$$P^+ = 2 \sum_n \left(k_n^+\right)^2 q_n q_{-n}\;.\eqno(2.10)$$

\vspace{.1in} \noindent One can show that $P^+$ has vanishing Dirac brackets
with $\Sigma_n$, $q_0$, and $P^-$:

$$\left[P^+,\Sigma_n\right] = \left[P^+, q_0\right] = \left[P^+,P^-\right] =
0\; .\eqno(2.11)$$\\

\begin{large} \begin{bf} \noindent III.  Canonical Quantization \end{bf}
\end{large}

\vspace{.1in}
To quantize the system one replaces the classical fields with corresponding
field operators. One uses commutators instead of Dirac brackets and inserts a
factor of $i$. One must choose a regularization and an operator-ordering
prescription in order to make the system well-defined.

We begin by defining creation and annihilation operators $a_k^\dagger$ and
$a_k$,
$$q_k = \sqrt {d\over{4 \pi \left| k \right|}} \: a_k\;,\quad a_k =
a_{-k}^\dagger\;,\quad  k\neq 0\; , \eqno(3.1)$$

\vspace{.1in}
\noindent which satisfy the usual commutation relations
$$\left[a_k,a_l\right] =0\;,\quad\left[a_k^\dagger,a_l^\dagger\right]
=0\;,\quad
\left[a_k,a_l^\dagger\right] =\delta_{k, l}\; ,\quad k,l\neq 0\; .\eqno(3.2)$$

\vspace{.1in}
\noindent We also define
$$q_0 = \sqrt{d\over{4 \pi}} \: a_0\;.\eqno(3.3)$$

\vspace{.1in}
Very general arguments suggest that the Hamiltonian should be symmetric ordered
\cite{benderpinsky}. This choice of operator ordering produces
tadpoles which we eliminate by adding an overall constant and a
mass counterterm to the
Hamiltonian:

$$-{{\mu^2 d}\over {8 \pi}}{\sum_{l>0} {1\over l}}+{{\lambda d}\over{128
\pi^2}}
\left\{\sum_{l>0}{1\over l}\right\}^2 - {{\lambda d}\over{64 \pi^2}} \left\{
\sum_{l>0}{1\over l}\right\} {\sum_k {{a_k a_{-k}}\over l}}\;. \eqno(3.4)$$

\vspace{.1in}
\noindent
Then, the quantum Hamiltonian is

\begin{eqnarray*} P^- &=& {{\mu^2 d a_0^2}\over {8 \pi}} + \mu^2 \Sigma_2 +
{{\lambda d}\over{4! 16 \pi^2}} {\sum_{k, l, m, n} {{a_k a_l a_m a_n \,
\delta_{k+l+m+n, 0}}\over{\sqrt{k l m n}}} }\\  & &+ {{\lambda d}\over{128
\pi^2}} \left(\sum_{l>0}{1\over l}\right)^2 - {{\lambda d}\over{64 \pi^2}}
\left(\sum_{l>0}{1\over l}\right) {\sum_k {{a_k a_{-k}}\over l}}\;.
\end{eqnarray*}
\vspace{-0.6in}$$\eqno(3.5)$$

\vspace{.1in}
\noindent
Note that the constraint equation for the zero-mode is obtained by taking a
derivative of $P^-$ with respect to $a_0$. Consequently, it is natural in
the quantum case to order symmetric order the constraint equation:
$$0 = g a_0 + a_0^3 + 2 a_0 \Sigma_2 + 2 \Sigma_2 a_0 + {\sum_{k > 0} {{a_k a_0
a_{-k} + a_{-k} a_0 a_k - a_0}\over k}} + 6 \Sigma_3\;,\eqno(3.6)$$

\vspace{.1in}\noindent
where $g= 24 \pi \mu^2/\lambda$. This equation implies that in matrix
form $a_0$ is real and symmetric. Moreover, it is block diagonal in states with
equal $P^+$ eigenvalues. Using the constraint equation, we define a rescaled
Hamiltonian $H$:

\goodbreak\begin{eqnarray*}
H ={{96 \pi^2}\over{\lambda d}} P^- &=& g \Sigma_2 +g\Sigma_4 -
{a_0^4\over 4} -{{a_0 \Sigma_2 a_0}\over 2}\\ & &+{1\over 4 }{\sum_{j, k, l
\neq
0}{{a_j a_0 a_k a_l +a_j a_k a_0 a_l}\over\sqrt{j k l}}\delta_{j+k+l,0}}\\ & &+
{1\over 4}{\sum_{k>0}{{a_k a_0^2 a_{-k} +a_{-k} a_0^2 a_{k} - a_0^2}\over
k}}\;.
\end{eqnarray*}\vspace{-0.62in} $$\eqno(3.7)$$\\

\begin{large} \begin{bf}\noindent
IV.  One Mode, Many Particles
\end{bf}\end{large}

\vspace{.1in}
It is reasonable to assume that the lowest-energy mode will be the most
important one in the constraint equation (3.6). We therefore study the case
where the zero-mode is just a function of the lowest-energy mode. In this case,
the zero-mode is diagonal and can be written as

$$a_0 = f_0 \left| 0 \right\rangle\left\langle 0 \right| + {\sum_{k>0}
f_{k} \left| k \right\rangle\left\langle k \right|}\;.\eqno(4.1)$$

\vspace{.1in}
\noindent
Equivalently, one can think of the zero-mode as an operator-valued function of
the number operator $N = a^\dagger a$. The VEV is given by
$$\left\langle 0 \right| \phi \left| 0 \right\rangle ={1\over{\sqrt {4 \pi}}}
\left\langle 0\right| a_0 \left| 0\right\rangle ={1\over{\sqrt {4 \pi}}}
f_0\;.\eqno(4.2)$$

\vspace{.1in}
Substituting (4.1) into the constraint equation (3.6) and sandwiching the
constraint equation between Fock states, we get a
recursion relation for $f_{n}$:

$$0 =  g f_{n} + {f_{n}}^{3} + (4n - 1) f_{n} +
\left(n+1\right) f_{n+1} + n f_{n-1}\;.\eqno(4.3)$$

\vspace{.1in}
\noindent If we take $f_0 =0$ and we assume that $n f_{n-1}$ evaluated at $n=0$
vanishes \cite{note}, then we see that all the $f_k$'s are zero. This
corresponds to the unbroken phase, $a_0 = 0$. Our objective is to determine
whether spontaneous symmetry breaking occurs and a nonzero solution for $a_0$
appears as we increase $\lambda$ (decrease $g$).

We begin our analysis of Eq.~(4.3) by finding its asymptotic behavior for large
$n$. If $f_{n}\gg 1$ in this limit, then the ${f_{n}}^3$ term will dominate and

$$f_{n+1} \sim {{f_{n}^3}\over n}\;, \eqno(4.4)$$

\vspace{.1in}
\noindent from which we deduce that

$${\lim_{n\to\infty} f_{n}} \sim (-1)^n \exp\!\left(3^n {\rm
constant}\right)\;. \eqno(4.5)$$

\vspace{.1in}
We now argue that we reject this rapidly growing solution. The part of the
Hamiltonian (3.7) corresponding to the lowest mode is diagonal in $N$ and the
zero-modes will affect all of the energy levels. If $f_{n}$ is large for large
$n$ then the high energy levels will be strongly affected. The paradigm for
spontaneous symmetry breaking is the symmetric double well which has a ground
state localized in either well rather than at the symmetry point. This paradigm
indicates that the behavior of the system is unaffected by the barrier for
energies far above the barrier separating the wells.
Hence, we only seek solutions where $f_{n}$ is small for large $n$. This is the
central condition that will be used to determine the critical couplings in all
the subsequent calculations. We therefore neglect the $f_{n}^3$ term for large
$n$; it is the terms linear in $f_{n}$ that dominate, giving

$$f_{n+1} + 4 f_{n} + f_{n-1} = 0\;. \eqno(4.6)$$

\vspace{.1in}
\noindent
There are two solutions to this equation:
$$f_{n} \propto \left(\sqrt{3} \pm 2\right)^{n}\;.\eqno(4.7)$$

\vspace{.1in} \noindent
We reject the plus solution because it grows with $n$.  Dropping the cubic
term from (4.6) we define the generating function

$$F\!\left(z\right) = \sum_{n=0}^{\infty} f_{n} z^n\;.\eqno(4.8)$$

\vspace{.1in} \noindent If $f_{n}$ goes like $\left( \sqrt{3} -2\right)^n$ then
the radius of convergence of $F\!\left( z\right)$ is $2+\sqrt{3}$ and we expect
$F\!\left(z\right)$ to be singular at $\left|z\right| = 2+\sqrt{3}$. Similarly,
if $f_{n} \sim\left(\sqrt{3} + 2\right)^n$, then we expect $F\!\left(z\right)$
to be singular at $\left|z\right| = 2 -\sqrt{3}$.

The function $F\!\left(z\right)$ satisfies the differential equation
$${1\over{F\!\left(z\right)}}{{d F\!\left(z\right)}\over{d z}} = - {{g-
1+z}\over{z^2+4 z+1}}\;,\eqno(4.9)$$

\vspace{.1in}
\noindent whose solution is

$${{F\!\left(z\right)}\over{F\!\left(0\right)}}=\left( {{z+2-\sqrt 3}\over {2
-\sqrt 3}}\right)^{-{{\sqrt 3 - 3 + g}\over{2\sqrt 3}}}  \left({{z+2+\sqrt 3}
\over{2 + \sqrt 3}}\right)^{- {{\sqrt 3 +3-g}\over{2\sqrt 3}}}\;.\eqno(4.10)$$

\vspace{.1in}
\noindent Note that this solution for $F\!\left(z\right)$ has singularities at
the expected values of $z$. If we want $f_{n}$ to have the asymptotic behavior
$\left(\sqrt{3} - 2 \right)^n$ for large $n$, then we must eliminate the branch
point of $F\!\left(z\right)$ at $\left|z\right| =2-\sqrt 3$. This gives the
condition
$$-{{\sqrt{3} - 3 + g}\over{2\sqrt 3}} = K\;,\quad K= 0, 1, 2\;. \eqno(4.11)$$

\vspace{.1in}
\noindent
Only $K = 0$ gives $g>0$. Therefore, we find a critical coupling
$$g_{\rm critical} = 3 - \sqrt{3}\;. \eqno(4.12)$$

\vspace{.1in}
\noindent We conclude that there is a critical value of the coupling constant
for which there is a nonzero value for $a_0$ and a solution to the linearized
equation for $f_{n}$ exists that does not grow rapidly for large $n$. We can
compare our critical value of $g$ with that obtained in the equal-time
formulation. Chang \cite{chang} finds that $\lambda_{\rm critical} = 54.3$
which differs from our result

$$\lambda_{\rm critical}=4\pi\left(3+\sqrt{3}\right)\approx 59.5\eqno(4.13)$$

\vspace{.1in}
\noindent
by about 10\%.

Of course, we need to determine if there is just an isolated critical point or
if there is a continuous range of values of $g < g_{\rm critical}$ for which
$a_0$ has nontrivial solutions. This requires that we investigate the full
nonlinear equation. Away from $a_0 =0$ where the $a_0^3$ term can make
substantial contributions we will use both the $\delta$-expansion and numerical
methods to answer this question. We denote the values of $f_0$ vs. $g$ that
satisfy (4.3) the ``critical curve''.\\

\begin{large}\begin{bf}\noindent
V.  The $\delta$-expansion
\end{bf}\end{large}

\vspace{.1in}
The $\delta$-expansion is a powerful perturbative technique for linearizing
nonlinear problems. It has been shown to be an accurate technique for solving
problems in differential equations, quantum mechanics, and quantum field theory
\cite{bendermilton}.

We rewrite Eq.~(4.3) as

$$\left(g - 1 + 4 n\right)f_{n} + f_{n}^{1+2\delta} + \left(n +
1\right) f_{n+1} + n f_{n-1} = 0. \eqno (5.1)$$

\vspace{.1in}
\noindent
Setting $\delta = 0$ gives the linear finite difference equation which
is the zeroth-order approximation in the $\delta$-expansion. One then expands
in
powers of $\delta$ about $\delta = 0$. One recovers the problem of interest
at $\delta = 1$. Expanding about $\delta = 0 $ we have

$$g = g^{\left(0\right)} + {\delta g}^{\left(1\right)} + \ldots \;, \quad
f_{n} = f_{n}^{\left(0\right)} + {\delta}f_{n}^{\left(1\right)} + \ldots \;,
\quad n=1, 2, \ldots \eqno(5.2)$$

\vspace{.1in}\noindent
and

\begin{eqnarray*} f_{n}^{1 + 2{\delta}} &=& f_{n}\left(1 + \delta \ln f_{n}^2 +
\ldots\right)\\ &=& f_{n}^{\left(0\right)}+\delta\!
\left(f_{n}^{\left(0\right)}
+f_{n}^{\left(0\right)}\ln {f_{n}^{\left(0\right)}}^2\right) + \ldots\;.
\end{eqnarray*}
\vspace{-0.6in}$$\eqno(5.3)$$

\vspace{.1in}\noindent
Substituting this into equation (4.1) we find, to zeroth order in $\delta$,

$$\left(g^{\left(0\right)} +4 n \right) f_{n}^{\left(0\right)} + \left(n
+1\right)f_{n}^{\left(0\right)} + n f_{n-1}^{\left(0\right)} = 0\;.
\eqno(5.4)$$

\vspace{.1in}\noindent
This zeroth-order equation is the same equation as that studied in
Section~IV with $g$ displaced by one. As discussed before, we can determine the
solution for large $n$. One then finds that
$$f_{n}^{\left(0\right)} = f_0\left(\sqrt 3
-2\right)^n\;,\;\;\;\;\;  g^{\left(0\right)} = 2-\sqrt 3\;.\eqno(5.5)$$

\vspace{.1in}
To first order in $\delta$ we obtain an inhomogeneous second-order finite
difference equation:
$$\left(g^{\left(1\right)} + 4 n\right) {f_{n}}^{\left(1\right)} +
\left(n+1\right) {f_{n+1}}^{\left(1\right)}+ n
{f_{n-1}}^{\left(1\right)} = -{f_{n}}^{\left(0\right)}\left(\ln {{f_{n}}^{
\left(0\right)}}^2 + g^{\left(1\right)} \right) \;.\eqno(5.6)$$

\vspace{.1in}
\noindent This equation can also be solved exactly. We obtain

$$f_{n}^{\left(1\right)} = -{{\left(\sqrt 3 - 2\right)^n f_{0}}\over{\sqrt 3}}
\ln\left(2 + \sqrt 3\right) + \left(\sqrt 3 - 2\right)^n \left[\ln f_{0}^2 +
g^{\left(1\right)} + {1\over{\sqrt 3}} {{\ln \left(2  - \sqrt{3}\right)}
\over{2 + \sqrt 3}}\right]$$
$$ \times \left\{\left(2+\sqrt 3\right)^2  \sum_{p=1}^n
{\left(2+\sqrt 3\right)^{2 p}\over p} - \sum_{p=1}^n {1\over p}\right\}
{f_{0}\over {2\sqrt 3}}. \eqno(5.7)$$

\vspace{.1in}
\noindent The second term in (5.7) grows with $n$ so we demand that its
coefficient vanishes, which gives
$$g^{\left(1\right)} = -\ln f_{0}^2 + {1\over{\sqrt 3}} \ln \left(2 +
\sqrt 3\right)\left(2 - \sqrt 3\right)\;, \eqno(5.8)$$
$${f_{n}}^{\left(1\right)} = - {{f_{0}}\over{\sqrt 3}}
\left(\sqrt 3 - 2\right)^n \ln \left( 2 + \sqrt 3\right)\;. \eqno(5.9)$$

\vspace{.1in} \noindent These results are plotted in Fig.~1. The dashed line is
a plot of the critical curve for $\delta = 1$, \\
$$g = \left(2-\sqrt 3\right) \left(1+{1\over{\sqrt 3}} \ln \left(2+\sqrt
3\right)\right) - \ln {f_{0}^2}\;,\eqno(5.10)$$

\vspace{.1in}
\noindent away from $f_{0} = 0$. The expansion behaves badly near $f_{0}=0$
because of the $\ln f_{0}^2$ term in the expansion. The $\delta$-expansion
analysis clearly shows that there is a critical curve and not merely a critical
point.\\

\begin{large} \begin{bf} \noindent VI.  Numerical Solution \end{bf} \end{large}

\vspace{.1in} We can study this critical curve in detail by looking for
numerical solutions to Eq.~(4.3). The method used here is to write Eq.~(4.3) as
a set of $M$ simultaneous equations:

\begin{eqnarray*}
0 &=& \left(g-1\right) f_{0} + f_{0}^3 + f_{1} \;,\\
0 &=& \left(g+3\right) f_{1} + f_{1}^3 + 2 f_{2} +   f_{0} \;,\\
0 &=& \left(g+7\right) f_{2} + f_{2}^3 + 3 f_{3} + 2 f_{1} \;,\\
 &\vdots& \\
0 &=&\left(g+4 M-1\right) f_M + {f_M}^3 + M f_{M-1} \; .
\end{eqnarray*}
\vspace{-0.6in}$$\eqno(6.1)$$

\vspace{.1in}
\noindent In the $M$th equation $f_{M+1}$ is set to zero. Since we seek a
solution where $f_{n}$ is decreasing with $n$, this is a good approximation. We
then pick a value of $g$ and look for real solutions for $f_{0}, f_{1}, \ldots,
f_{M}$. We find that for $g>3-\sqrt 3$ the only real solution is $f_{n}=0$ for
all $n$. For $g$ less than $3-\sqrt{3}$ there are two additional solutions and
near
the critical point $\left| f_{0} \right|$ is small and

$$f_{n} \approx f_{0} \left(2 - \sqrt 3\right)^n  \eqno(6.2)$$

\vspace{.1in}\noindent
As $g$ decreases ($\lambda$ increases), the solution for $\left| f_{0}
\right|$ increases. The critical curve is indicated by the solid line in
Fig.~1.
The solution of (6.1) converges quite rapidly with $M$. The critical curve is
approximately parabolic in shape:

$$g =  3-\sqrt 3 - 0.9177 f_{0}^2  \eqno(6.3)$$

\vspace{.1in}\noindent
For a given value of $f_{0}$ and $g$ Eq.~(4.2) can be used to
calculate all values of $f_{n}$.

It is interesting to study the behavior of the constraint equation (4.3) away
from the critical curve. In Fig.~2 we plot $\left| f_{n} \right|$ as a function
of $n$ and $f_{0}$ for $g = 1.2$. We see that, as $n$ becomes large, all the
$\left| f_{n} \right|$ increase and as $f_{0}$ approaches the critical curve,
which is at $f_{0} \approx 0.2700$ for $g=1.2$, all the $\left| f_{n}\right|$'s
decrease rapidly. As $f_{0}$ increases beyond the critical curve the $\left|
f_{n}\right|$'s increase rapidly once again. The fact that $\left|
f_{n}\right|$
increases rapidly on both sides of the critical curve is a manifestation of the
nonlinearity in (4.3).

We can also study the eigenvalues of the Hamiltonian (3.7) for this one-mode
problem. The Hamiltonian is diagonal in the number operator $N$ so the energy
eigenstates are just the eigenstates of $N$.  Thus,

$$\left\langle n \right| H \left| n \right\rangle = {3\over 2} n (n-1) + n g
-{f_n^4\over 4} - {{2 n+1}\over 4} f_n^2 +{{n+1}\over 4} f_{n+1}^2 + {n\over 4}
f_{n-1}^2\;. \eqno(6.4)$$

\vspace{.1in}\noindent
In Fig.~3, the dashed lines show
the first few eigenvalues as a function of $g$ without the zero-mode. Observe
that the vacuum is at zero for all $g$.  When we include the zero-mode,
the energy levels shift as shown by the solid curves.  The vacuum is at zero
energy for $g > g_{\rm critical}$ but at $g = g_{\rm critical}$ there is a
phase
transition and the energy decreases below zero as g is decreased. We also see
that for $g < g_{\rm critical}$ all the higher energy level increase above the
value they had without the zero-mode. The higher levels change very little, as
our paradigm would suggest, because $f_n$ is small for large $n$.\\

\begin{large}\begin{bf}\noindent
VII.  Discussion
\end{bf}\end{large}

\vspace{.1in}
In the context of $\phi^4_{1+1}$ field theory on the light front, our paradigm
for spontaneous symmetry breaking suggests that spontaneous symmetry breaking
occurs when fields can develop a zero-mode. This zero-mode gives rise to a
nonzero VEV and the full vacuum remains the perturbative vacuum. In the broken
phase, the theory behaves exactly as expected and the numerical value of the
critical coupling $\lambda_{\rm critical} =4 \pi\left(3+\sqrt 3\right)$ agrees
well with the value obtained in the equal-time theory
$\lambda_{\rm critical} = 54.27$ \cite{chang,hari}.

Spontaneous symmetry breaking is simpler to understand in light-front field
theory because the entire effect comes from one mode. However, the problem of
solving for this one mode is quite difficult. In the literature, it has been
suggested that a direct solution of the zero-mode problem may be intractable
\cite{burkardt}. We hope to have convinced the reader that this is not true.

We interpret the existence of more than one solution to the constraint equation
to be spontaneous symmetry breaking. However, this spontaneous symmetry
breaking
is {\em non-dynamical}. We have no motivation to choose one solution to the
constraint equation over the others. The conventional argument that when the
system is coupled to a heat bath it will dynamically pick the lowest energy
state to be the vacuum does not apply here.

Many problems remain to be addressed. A more complete solution of the zero-mode
problem including all the oscillators would be very interesting, if only to
confirm our results. In reference \cite{heinzl}, the authors retain the one
particle and, implicitly, some of the two particle states for all of the modes
and find a
solution for the critical coupling $\lambda_{\rm critical} = 4 \pi \left(3.184
\ldots\right)$. Also, some questions regarding operator ordering still remain.
When
we chose a quantum Hamiltonian, we demanded that it be symmetrically ordered
and
we treated the zero-mode as an ordinary field operator. However, $q_0$ is not
an
ordinary field operator and can, in principle, be written in terms of the other
field operators $a_0 = c_0 +{\sum c_k a_k^\dagger a_k}+\ldots$. In this sense,
the Hamiltonian we chose is not really symmetrically ordered after all. It is
unclear whether this is a problem.\\

\newpage\begin{large}\begin{bf}\noindent
Figure Captions
\end{bf}\end{large}

\vspace{.1in}
\begin{description}
\item Figure 1. $g = 24 \pi \mu^2/\lambda$ vs. $f_0 = {\sqrt {4 \pi}} {\rm
VEV}$. The solid curve is the critical curve obtained from numerical solution
of (6.1) with $M=10$. The dashed curve is the critical curve obtained from the
first-order $\delta$-expansion.
\item Figure 2. $\left|f_n\right|$ as a function of $n$ and $f_0$ for $g=1.2$
from the numerical solution of (6.1) with $M=10$.
\item Figure 3. The lowest three energy eigenvalues as a function of $g$ from
the numerical solution of (6.1) with $M=10$. The dashed line is the symmetric
solution $f_0 = 0$ and the solid line is the solution with $f_0 \neq 0$ for
$g<g_{\rm critical}$.
\end{description}

\begin{thebibliography}{99}
\bibitem{note} A more general assumption would be to take $\lim_{n\to 0} n f_{n
-1}$ to be nonzero, introducing an additional arbitrary constant. In general
one
might expect a second order finite difference equation to require two boundary
conditions.
\end{thebibliography}
\end{document}